\documentclass[aps,prd,showpacs,nofootinbib,notitlepage, 12pt]{revtex4-1}
\usepackage{latexsym}
\usepackage{amsbsy,amsmath,amsthm,amsfonts,calc,amssymb,graphicx,accents,mathrsfs,eufrak}
\usepackage{braket,textcomp,upgreek}
\usepackage[percent]{overpic}
\usepackage{psfrag}
\usepackage{graphicx}
\usepackage{caption,subcaption}
\usepackage{color}
\usepackage{psfrag}
\usepackage{enumerate}
\usepackage[colorlinks,citecolor=blue]{hyperref}
\theoremstyle{plain}

\theoremstyle{definition}
\DeclareFontFamily{U}{rsfs}{}         
\DeclareFontShape{U}{rsfs}{m}{n}{<5> rsfs5 <6><7> rsfs7          %
 <8><9><10><10.95><12><14.4><17.28><20.74><24.88> rsfs10}{}     %
\DeclareMathAlphabet{\mathfs}{U}{rsfs}{m}{n}                     %
\definecolor{indiagreen}{rgb}{0.07, 0.53, 0.03}
\def\beq{\begin{eqnarray}}
\def\eeq{\end{eqnarray}}

\def\nn{\nonumber\\}

\def\={\stackrel{\Delta}{=}}

\def\lie{\pounds}

\DeclareMathOperator\erf{Erf}
\newcommand*\Laplace{\mathop{}\!\mathbin\bigtriangleup}

\begin{document}
\begin{flushright}
	\footnotesize USTC-ICTS/PCFT-20-31
\end{flushright}
\title{Dynamical horizons and Super-Translation transitions of the horizon} 
\author{Ayan Chatterjee}\email{ayan.theory@gmail.com}
\affiliation{Department of Physics and Astronomical Science, Central University 
	of Himachal Pradesh, Dharamshala -176215, India.}
\author{Avirup Ghosh}\email{avirup@ustc.edu.cn}
\affiliation{Interdisciplinary Center for Theoretical Study, University of Science and Technology of China, and 
	Peng Huanwu Center for Fundamental Theory, Hefei, Anhui 230026, China	}
\begin{abstract}
	  A condition is defined which determines if a supertranslation is induced in the course of a general evolution from one isolated horizon phase to another via a dynamical horizon. This condition fixes preferred slices on an isolated horizon and is preserved along an Isolated Horizon. 
	 If it is not preserved, in the course of a general evolution, then a supertranslation will be said to have been induced. 
	 A simple example of spherically symmetric dynamical horizons is studied to illustrate the conditions 
	 for inducing supertranslations.
\end{abstract}
\maketitle
\section{Introduction}
The notion of supertranslation and superrotation symmetries at the past and future null infinity have gained importance with the realisation that they reproduce Weinberg's soft graviton theorem \cite{He:2014laa,Kapec:2015vwa,Campiglia:2014yka,Campiglia:2015yka,PhysRev.140.B516} These fascinating results along with the proposal for resolving the information loss paradox by utilising the notion of supertranslations and interpreting them as additional hair for black holes \cite{Hawking:2016sgy,Haco:2018ske}, led to a renewed interest in the study of near horizon symmetries \cite{Donnay:2016ejv,Donnay:2015abr,Averin:2016ybl,Afshar:2016wfy,Setare:2016jba,Mao:2016pwq,Ciambelli:2019lap,Carlip:2017xne, Sousa:2017auc}. The main aim has been to explore the near horizon structure of black hole horizons and seek symmetries which are similar in spirit to the asymptotic symmetries at null infinity. While it is far from clear whether these indeed resolve the paradox it is also important that we explore these from a wider perspective.

These approaches have been directed towards stationary event horizons or quasi-local horizons and deal with finding conserved charges corresponding to these symmetries. It has been found that with the strictest of boundary conditions, only supertranslation symmetries are admitted. There can be a further enhancement following a weakening of the boundary conditions, but it is not completely understood whether such a weakening is indeed physical. There is however no confusion regarding the existence of supertranslation freedom.  These charges howver turn out to be trivial except for the zero mode, which gives the energy of the isolated horizon. This seems to indicate that these are purely gauge and are of no physical significance. While it is true that gauge symmetries will yield trivial Noether charges, the converse may not be true. Hence it is important that supertranslation freedom is explored from different perspectives. One proposition would therefore be to study these in the dynamical regime. This is analogous to studying how the structures at null infinity change when some physical process takes place in the bulk and consequently some radiation passes through null infinity \cite{Compere:2018ylh}. These include the memory effect where due to passage radiation through null infinity, the natural frames in the two stationary epochs, before and after the radiation has passed, are related by a supertranslation and a boost \cite{Strominger:2014pwa, Hollands:2016oma}. The memory effect can also be interpreted as a process where a supertranslation is being induced. Incidentally these effects are reflected in the evolution of the supertranslation and super-rotation charges at null infinity. 

There are however many problems that accompany this kind of an approach for black hole horizons. The first being that supertranslation is not a symmetry in the dynamical regime, thus raising doubts as to what the the expression for the charges would be during the evolution process. The second being the fact that unlike asymptotic null infinity the zeroth order structure at the horizon is not universal and in fact becomes dynamic during the evolution process thus posing as a hurdle in the way. A simplistic approach to studying the process of inducing a supertranslation to quasi-local black holes was taken in \cite{Ghosh:2020wjx}. Here we will go a step forward and study general dynamical evolutions and demonstrate how they may be induced for spherically symmetric dynamical horizon. As a consequence we will discuss some qualitative features of the infalling flux required so that a supertranslation is induced. 

In order to do so we first need to define what we would mean when we say that a supertranslation has been induced. It is known that in the case of Isolated horizons (IH's) the supertranslation freedom can be utilised to choose preferred slices of the horizon, the so-called good cuts. There are various choices for such good cuts \cite{Rahman:2019bmk}. These are reperesented as scalar conditions on the cross-sections of the horizon. It can be a condition on the connection in the normal bundle spanned by the two null vectors or can be a condition on the expansion in the transverse direction. In the case of an IH such a condition is preserved as time evolves. This howver may not be the case for Dynamical Horizons (DH's). The first aim here will be to find whether such a condition is preserved during a dynamical evolution.

In this case we define preferred slices as in \cite{Ashtekar:2001is}, by demanding that that the divergence of the so called rotation one form is zero. We find general evolution equations for this quantity and show that in general it is non on a DH. There is howver a choice of the shift vector, on a DH, such that the derivative of the divergence can be made to vanish. It is already clear that one can rescale the null vectors in the normal bundle to achieve the divergence free condition. Our results try to decipher what kind of evolution vector does so. But in doing so one induces an extra diffeomorphism on the cross sections (due to modification of the shift vector). Since diffeomorphisms on the cross section are not symmetries in IH phase \cite{Ghosh:2020wjx}, the final black holes formed in each case are indeed different. Symmetry requirements can however put severe constraints on the shift vector. In such cases there may not be enough freedom in the choice of the shift vector. This is demonstrated for the case where the dynamical horizon is spherically symmetric and the form of the stress energy tensor required to support such a process is partially constructed. 

The notations that will be used are as follows. The null vectors in spacetime will be denoted by $l,~n$. The coordinates on any two-surface will be denoted by $\tau^A$ while the coordinate vectors spanning the two surfaces will be denoted by $\partial_A$. The metric on any such two surface will be denoted by $q$. One forms will be denoted with an underline e.g ~$\underline{n}$.

\section{Action of super-translations on the isolated horizon data}
Following the definition of an isolated horizon (IH) \cite{Ashtekar:1998sp,Ashtekar:1999yj,Ashtekar:2000sz,Ashtekar:2000hw,Ashtekar:2001is,Ashtekar:2001jb} the two geometric quantities that transform under a supertranslation are the pull back of the rotation one form viz. $\omega_A:=-g(n,\nabla_{\partial_A}l)$ and the transverse extrinsic curvature $K^{(n)}_{AB}:=g(n,\nabla_{\partial_A}\partial_B)$ \cite{Ghosh:2020wjx}. Let us denote an IH by $\Delta$. A supertranslations can be viewed as a map $\psi:\tilde\Delta\rightarrow\Delta$. The transformation of $\omega_A$ can be obtained by considering the pull-back connection under the map $\psi$. The definition of the pull-back connection is,
\begin{gather}
\psi_*\bigg((\psi^*\mathbb D)_XY\bigg)=\mathbb D_{\psi_*X}\psi_*Y,
\end{gather}
where $\mathbb D$ is the connection on $\Delta$ and $X,~Y$ are arbitrary vectors on $\Delta$. The  basis vectors on $\Delta$ transform as $\psi_*l=l$, $\psi_*\partial_A=\partial_A\mathcal F~l+\partial_A$, while the one form $n$ transforms as $\psi^*n=n-\kappa ~d\mathcal F$. $\mathcal F$ is some function on the cross section $S^2$ of $\Delta$. Thus we have,
\begin{gather}
\psi_*(\tilde\kappa~l)=\kappa~ l\\
\psi_*(\tilde\omega_A ~l)=(\omega_A+\kappa~\partial_A\mathcal F)l.
\end{gather}
For the case of the extrinsic curvature $K^{(n)}(\partial_A,\partial_B)$, we have,
\begin{gather}
\psi_*\bigg((\psi^*\mathbb D)_{\partial_A}\partial_B\bigg)=\mathbb D_{\psi_*\partial_A}\psi_*\partial_B=\mathbb D_{\partial_A}\partial_B+\mathcal D_A\mathcal F~\mathbb D_{\partial_B}l+\mathcal D_B\mathcal F~\mathbb D_{\partial_A}l+\mathcal D_A\mathcal F~\mathcal D_B\mathcal F~\mathbb D_l~l\nn
+\mathcal D_A\mathcal D_B\mathcal F~l,
\end{gather}
where the covariant derivative compatible with the metric on the cross section, $q$, has been denoted by $\mathcal D$. On contracting both sides by $n-\kappa~d\mathcal F$ we get the desired result found in \cite{Ghosh:2020wjx}.
\section{Fixing a foliation}
In this section we would explore how a preferred foliation may be chosen by choosing $\omega_A$ to be divergence free. Note that any one form on the two sphere can be decomposed into a divergence and a divergence free part viz.
\begin{gather}
\omega_A=\Omega_A+\partial_A\omega\label{decompomega}
\end{gather}
where $\mathcal D_A\Omega^A=0$ and $\omega$ may be determined by the following expression
\begin{gather}
\omega(\tau):=\int G^{S^2}(\tau,\tau')\mathcal D^A\omega_A(\tau')d\tau',
\end{gather}
where $G^{S^2}(\tau,\tau')$ is the Green's function on the sphere. By choosing a particular foliation one can set $\mathcal D^A\omega_A$ and thus $\omega$ to zero on each cross-section \cite{Ashtekar:2001is}. In the case of an IH this is possible by application of a supertranslation. The dominant energy condition then ensures that this preferred choice of foliation is preserved along the IH, that is preferred leaves are mapped to preferred ones by the evolution vector. There are other  preferred choices for foliation \cite{Rahman:2019bmk} that one may also consider.

Note that in an IH phase, $\mathcal D^A\omega_A$ is preserved in time and thus fixing a particular leaf uniquely fixes the other leaves. This might not be the case during the course of a dynamical evolution. We will say that during a dynamical process a supertranslation has been induced if the leaves of the final IH formed does not satisfy the condition $\mathcal D^A\omega_A=0$ even though the leaves of the initial IH did. Our main aim would be to check whether there exists an evolution vector that indeed preserves the divergence, if not to seek insights into the content of the ingoing flux that induces such a change.
\section{Evolution equations}
We will now try to evaluate the evolution of the divergence along a Dynamical horizon. For that purpose let us recall the definition of a dynamical horizon (DH)\cite{Ashtekar:2003hk}. A dynamical horizon $\mathcal H$ is defined by a spacelike surface foliated by marginally trapped surfaces (MTS's), given by the conditions $K^{(l)}=0$ ($l$ is outgoing and future directed null normal and $K$ denotes the trace of the extrinsic curvature). Thus each cross-section $S_{\mathcal H}$ of $\mathcal H$ is a MTS. Let us assume that that the null frame in the equilibrium regions have been smoothly extended  so as to construct a null frame on each of the $S_{\mathcal H}$'s. The evolution vector which maps one $S_{\mathcal H}$ to another can then be taken to  $X^\perp=\alpha~l-\beta~n$ \cite{Booth:2006bn}. A timelike vector $\tau=\alpha~l+\beta~n$, orthogonal to $X^\perp$ can also be constructed.  The condition that $X^\perp$ mpas one MTS to another gives a constraint $\alpha$ and $\beta$. This is a partial differential equation on $S_{\mathcal H}$,
\begin{gather}
\Laplace^{S_{\mathcal H}}\beta-2\omega^A\partial_A\beta-\beta \mathcal D_A\omega^A+\beta\omega^A\omega_A-\beta\bigg(\frac{^2\mathcal R}{2}-8\pi G~T(l,n)\bigg)\nn
-\alpha\bigg(8\pi G~T(l,l)+K^{(l)}_{AB}K^{(l) AB}\bigg)=0,\label{MOTS}.
\end{gather}
In the above equation the Laplace operator and the covariant derivative on $S_{\mathcal H}$ has been denoted by $\Laplace^{S_{\mathcal H}}$ and $\mathcal D$ respectively. The the extrinsic curvature of $l$ has been denoted by $K^{(l)}_{AB}$. $T(X,Y)$ is the stress energy tensor and $^2\mathcal R$ is the Ricci scalar on $S_{\mathcal H}$. This expression can also be found in \cite{Booth:2006bn} e.g. The change in the area of the cross-section $S_{\mathcal H}$ is given by,
\begin{equation}
\lie_X~^2\epsilon=\bigg(\alpha K^{(l)}-\beta K^{(n)}\bigg)~^2\epsilon,
\end{equation}
In order that divergence free vectors are mapped to divergence free vectors (which are used to define horizon multipole moments \cite{Ashtekar:2004gp,Ashtekar:2013qta}) one has to add a component which is tangential to the cross sections $S_{\mathcal H}$. Thus we modify $X^{\perp}$ as $X=X^\perp+N^T$, such that \cite{Ashtekar:2013qta}
\begin{gather}
div (N^T)=\frac{2\dot R}{R}+\beta K^{(n)}
\end{gather}
where $R$ is the areal radius defined as $R^2:=\frac{1}{4\pi}\int \sqrt{q}~d^2\tau$. Note that the above condition only fixes the divergence part of the shift vector $N^T$. It however does not specify the curl or divergence free part of the vector field. We will see what use this part is of. Before discussing the evolution of the geometric structures, let us see how the supertranslation charges evolve. This will indirectly be of relevance in our latter discussions. Recall that in \cite{Ashtekar:2003hk} the surface gravity $\kappa$ was defined as $1/2R$. This is the form that is consistent with a local differential form of the first law. Here we will assume that in the dynamical phase the surface gravity is given by, $\kappa:=-g(n,\nabla_{X^\perp}l):=1/2R$. This essentially fixes a gauge freedom. A generalisation of the supertranslation charges given in \cite{Ghosh:2020wjx} to the dynamical phase may be written as,
\begin{gather}
Q_{f}:=\frac{1}{4\pi G}\int_{S^2}\kappa f~~^2\epsilon=\frac{1}{8\pi GR}\int_{S^2}f~~^2\epsilon.
\end{gather}
With the assumed expression for the surface gravity we have,
\begin{gather}
\dot{Q_f}=-\frac{\dot R}{8\pi GR^2}\int_{S^2}f~~^2\epsilon+\frac{2\dot R}{8\pi GR^2}\int_{S^2}f~~^2\epsilon=\frac{\dot R}{R}Q_f\label{evos}
\end{gather}
The above equation implies that if the a supertranslation charge is zero for the initial black hole, then it remains so for the final black hole formed. This is the consequence of the demand that a local differential first law holds.
\subsection{Supertranslations}
As has been discussed before a non trivial evolution of the divergence of the rotation one form will indicate that a supertranslation is being induced. We therefore require an expression for the evolution the divergence of the rotation one -form. Note that $\omega_A$ is a number in spacetime but a one form under coordinate transformations on $S_{\mathcal H}$. In order to find its variation along $X$ it is sufficient to take its covariant derivative along $X$,
\begin{gather}
-\nabla_{X}\omega_A
=g(\nabla_Xn,\nabla_{\partial_A}l)+g(n,R(X,\partial_A)l~)+\nabla_{\partial_A}g(n,\nabla_X l)
-g(\nabla_{\partial_A}n,\nabla_X l)\nn
\nn
=K^{(n)}(X^T,\partial_C) q^{CD}K^{(l)}_{DA}-K^{(l)}(X^T,\partial_C) q^{CD}K^{(n)}_{DA}-(\partial_C\alpha+\alpha\omega_C) q^{CD}K^{(l)}_{DA}\nn
-(\partial_C\beta-\beta\omega_C) q^{CD}K^{(n)}_{DA}+g(n,R(X,\partial_A)l~)-\partial_A\kappa_X,\label{evo1}
\end{gather}
where an expansion of the normal part of the evolution vector as $\alpha~l-\beta~n$ has been done. $\kappa_X$ here denotes $-g(n,\nabla_Xl).$
Let us now try to write down the Riemann tensor in terms of quantities realisable through Einsteins' equations. Using the Codazzi equation one can obtain the following expression for the component of the Riemann tensor in question. 
\begin{gather}
g(n,R(X,\partial_A)l~)=-R(\tau,\partial_A)+\alpha\bigg(\mathcal D _AK^{(l)}-\mathcal D^BK^{(l)}_{BA}+\omega ^BK^{(l)}_{BC}-\omega _AK^{(l)}\bigg)\nn
+\beta\bigg(\mathcal D _AK^{(n)}-\mathcal D ^BK^{(n)}_{BA}-\omega ^BK^{(n)}_{BC}+\omega _AK^{(n)}\bigg)+g(n,C(X^T,\partial_A)l~),
\end{gather}
 Using this expression in eq. (\ref{evo1}) one arrives at the following expression.
 \begin{gather}
 \nabla_X\omega^A=q^{AB}\bigg[-K^{(n)}(X^T,\partial_C) q^{CD}K^{(l)}_{DB}+K^{(l)}(X^T,\partial_C) q^{CD}K^{(n)}_{DB}+\mathcal D_C\big(q^{CD}K^{X^\perp}_{DB}\big)+R(\tau,\partial_B)\nn
 -\alpha \mathcal D_{\partial_B}K^{(l)}-\beta \mathcal D_{\partial_B}K^{(n)}+\omega_B~K^{X^\perp}+\partial_B\kappa_X+g(n,C(X^T,\partial_B)l~)\bigg]\nn
 -\omega_B q^{AC}q^{BD}\bigg[g(\nabla_{\partial_C}N^T,\partial_D)+g(\nabla_{\partial_D}N^T,\partial_C)-2K^{X^\perp}_{CD}\bigg]\label{omegaev}
 \end{gather}
 The evolution of the divergence is just the divergence of the right hand side of eq. (\ref{omegaev}) because $X$ maps divergence free vectors to divergence free vectors. Let us try to make sense of this equation by decomposing some of these terms into irreducible parts. In particular we would be interested in terms which contain the curl or divergence free part of the vector field $N^T$. Let us consider the first term. Note that only the trace free part $\Sigma_{AB}$ of $K_{AB}$ contributes to the first two terms. Moreover this is an antisymmetric tensor on a two surface and therefore must be proportional to the area two form. Hence we have,
 \begin{gather}
 -K^{(n)}(X^T,\partial_C) q^{CD}K^{(l)}_{DB}+K^{(l)}(X^T,\partial_C) q^{CD}K^{(n)}_{DB}=\alpha~(N^T)^A~\epsilon_{AB}\label{1stterm}
 \end{gather}
 On decomposing $N^T$ as $(N^T)^A=\epsilon^{AB}\partial_Bg+q^{AB}\partial_Bf$ the condition on the divergence reduces to a condition on $f$ viz. $\mathcal D^2f=\frac{2\dot R}{R}+\beta K^{(n)}$. The two constants in the solution of this equation can be fixed via the condition $\int f~\sqrt{q}~d^2\tau=1$ \cite{Ashtekar:2013qta}. Putting this into the expression eq. (\ref{1stterm}) and taking the divergence yields,
 \begin{gather}
 -K^{(n)}(X^T,\partial_C) q^{CD}K^{(l)}_{DB}+K^{(l)}(X^T,\partial_C) q^{CD}K^{(n)}_{DB}=\mathcal D_B\alpha\bigg(q^{BC}\partial_Cg+\epsilon^{BC}\partial_Cf\bigg)+\alpha\mathcal D^2g
 \end{gather}
 Now let us consider the terms coming from $\nabla_{N^T}q^{AB}$. On rewriting these in terms of coordinates we have,
 \begin{gather}
-\omega_B q^{AC}q^{BD}\bigg[g(\nabla_{\partial_C}N^T,\partial_D)+g(\nabla_{\partial_D}N^T,\partial_C)\bigg]=-\omega_B q^{AC}q^{BD}\bigg[\mathcal D_CN^T_D+\mathcal D_DN^T_C\bigg]\nn
=-\mathcal D^A(\omega.N^T)+(\mathcal D^A\omega_B)(N^T)^B-\omega_B\mathcal D^B(N^T)^A
 \end{gather}
First, note that the first term cancels with the the $\partial_B\kappa_{N^T}$ term in $\kappa_X$. Let us consider the other terms. Noting that $\Omega$ in eq. (\ref{decompomega}) can be written as, $\Omega_A=\epsilon_A^{~B}\partial_B\Omega$, we have
\begin{gather}
(\mathcal D^2\omega_B)(N^T)^B+(\mathcal D_A\omega_B-\mathcal D_B\omega_A)\bigg(\mathcal D^A(N^T)^B\bigg)-\omega^B\mathcal D_B(div~ N^T)-\frac{~^2\mathcal R}{2} ~\omega_A(N^T)^A\nn
=\bigg(\mathcal D^2\omega_B-\frac{~^2\mathcal R}{2} ~\omega_B\bigg)\bigg(\epsilon^{BC}\partial_Cg+q^{BC}\partial_Cf\bigg)+2\mathcal D^A\mathcal D^B\Omega~\mathcal D_A\mathcal D_Bg-\omega^B\mathcal D_B\mathcal D^2f
\end{gather}
The main aim here would be write down an equation for $g$ such that the divergence of $\omega$ is preserved. We will therefore equate the divergence of the right hand side of eq. (\ref{omegaev}) to zero and find out an equation for $g$. The equation will clearly be of the form,
\begin{gather}
\bigg(\alpha q^{AB}+2\mathcal D^A\mathcal D^B\Omega\bigg)~\mathcal D_A\mathcal D_Bg+C^A\partial_Ag=\mathcal B\label{constraintangenet},
\end{gather}
where $C^A=(\mathcal D^2\omega_B-\frac{~^2\mathcal R}{2} ~\omega_B)\epsilon^{BA}+q^{BA} \mathcal D_B\alpha$ and $\mathcal B$ is some function on the cross section obtained from the rest of the terms. This term is known from the data and the function $f$ found previously. The existence of global solutions to this equation depends on the nature of the equation, parabolic, hyperbolic or elliptic. We will not deal with it here but assume that it does admit a global solution on $S^2$. In such a  case there exist a foliation of a DH such that no supertranslation is induced. But this does come at the cost of inducing an extra diffeomorphism on $S^2$ generated by the divergence free part of $N^T$. Since diffeomorphisms on the cross sections are not symmetries of an IH it seems that the black holes thus formed with and without this modification of $N^T$ are not related by some symmetry transformation. It is not surprising to see that one needs to implement a diffeomorphism on the cross section in order to avoid a supertranslation. This can clearly be seen by verifying that the phase space conjugate of a given supertranslation on the IH phase space is in fact a diffeomorphism on $S^2$ cross-sections. The modification however completely fixes the Shift vector, as opposed to previous consideration which only fixed the divergence part of the Shift.

In the next section we will try to give a more precise meaning of the action of diffeomorphisms on $S^2$ during a dynamical evolution. Before that let us check if the addition of the divergence free part to $N^T$ has any effect on the evolution of the multipole moments.
\begin{gather}
\lie_ X\int\xi^A\omega_A\sqrt{q}~d^2\tau=\int\bigg[\alpha~\xi^A(N^T)^B\epsilon_{AB}+\xi^AD_C\big(q^{CD}K^{X^\perp}_{DA}\big)\nn
+\xi^AR(\tau,\partial_A)+\beta\xi^A\partial_AK^{(n)}
+g(n,C(N^T,\xi)l)\bigg]
\end{gather}
On decomposing $\xi^A=\epsilon^{AB}\partial_Bh$, since it is divergence free, one can check that the above expression is not completely independent of $g$. So there might as well be a non trivial signature of the the modification in the evolution of the multipole moments.
 \subsection{Diffeomorphisms on the $\mathcal S^2$}
 Any coordinate choice on the cross-section can be written as a vector valued function of the connection and the metric $F^A(\gamma,q)$. To determine whether there is a diffeomorphism on $\mathcal S^2$ during the evolution, one needs to find whether this condition is preserved. This can only be done for known choices. For example if harminic coordinates could on $\mathcal S^2$ then it continues to be so, in the course of evolutions as,
 \begin{gather}
 \lie_X\mathcal D^2\tau^A=0.
 \end{gather}
 Thus there is no diffeomorphism on $\mathcal S^2$. There can be also be gauge choices which are expressed differently from the one mentioned above. In the case of axially symmetric DH the choice is the following. Let $\varphi$ be the vector field generating axis symmetry. Define a coordinate which is an affine parameter of $\varphi $ and normalized so to a length of $2\pi$.
 \begin{gather}
 \lie_{\varphi}\phi=1
 \end{gather}
 The other coordinate $\zeta$ is chosen such that,
 \begin{gather}
 \partial_B\zeta=\frac{1}{R^2}\varphi^A\epsilon_{AB}
 \end{gather}
 where $R$ is the area radius. The evolution vector is so chosen that it commutes with the Killing vector. Thus we have,
 \begin{gather}
 \lie_X\lie_\varphi\phi=\lie_\varphi\lie_X\phi=0,\nn
 \lie_X\partial_A\zeta-\lie_X\bigg(\frac{1}{R^2}\varphi^A\epsilon_{AB}\bigg)=0
 \end{gather}
One must also ensure that $\lie_\varphi~^2 q=0$ and $\lie_{\varphi}\bigg(\lie_X~^2q\bigg)=0$. This partially fixes the tangent component $X^T$. Thus there is not enough freedom in the choice of $X^T$ such that the previous condition eq. (\ref{constraintangenet}) be satisfied as well. Thus in every such case it might not be possible to keep the divergence of the rotation one form equal to zero. In order to understand the evolution in such a case we will consider collapse process where the DH is spherically symmetric, both from an analytic as well as numerical perspective.  
  \subsection{Evolution of transverse extrinsic curvature}\label{EVTE}
In this section we will give an expression for the evolution of the transverse extrinsic curvature $g(\tau,K(\partial_A,\partial_B))$. This will in general be useful for finding the content of the flux required to induce a supertranslation. In the next section while dealing with the spherically symmetric dynamical horizon this will be  a further consistency check for the flux required to induce a supertranslation. Recall that,
 \begin{gather}
 \nabla_XK(\partial_A,\partial_B)=\bigg(R(X^\perp,\partial_A)\partial_B\bigg)^\perp+\nabla^\perp_{\partial_A}\nabla^\perp_{\partial_B}X^\perp-\nabla^\perp_{\big(\nabla_{\partial_A}\partial_B\big)^T} X^\perp\nn
 -K\big(\partial_A,W_{X^\perp}(\partial_B)\big)-g\bigg(K(\partial_A,\partial_B),\nabla^\perp_{\partial_C}X^\perp\bigg)q^{CD}\partial_D\nn
 +\nabla_{N^T}K(\partial_A,\partial_B)-K([N^T,\partial_A],\partial_B)-K(\partial_A,[N^T,\partial_B]),
 \end{gather}
Now, using the fact that $\lie_Xg(\tau,K_{AB})=g(\nabla_{X^\perp}\tau,K_{AB})+g(\tau,\nabla_{X^\perp}K_{AB})+\lie_{N^T}K^{(\tau)}_{AB}$ and the above expressionexpression of the derivative of the extrinsic curvature, we have the following expression for the evolution of the transverse extrinsic curvature,
 \begin{gather}
 \lie_Xg(\tau,K_{AB})=\frac{1}{2}\bigg(2\kappa_{X^\perp}+\frac{\nabla_{X^\perp}\alpha}{\alpha}-\frac{\nabla_{X^\perp}\beta}{\beta}\bigg)K_{AB}^{X^\perp}+\frac{1}{2}\bigg(\frac{\nabla_{X^\perp}\alpha}{\alpha}+\frac{\nabla_{X^\perp}\beta}{\beta}\bigg)K_{AB}^{\tau}\nn
 +\alpha\mathcal D_A\mathcal D_B\beta-\beta\mathcal D_A\mathcal D_B\alpha-2\partial_{(A}(\alpha\beta)~\omega_{B)}-2\alpha\beta\mathcal D_A\omega_B
  -\alpha^2K^{(l)}_{AC}q^{CD}K^{(l)}_{BD}+\beta^2K^{(n)}_{AC}q^{CD}K^{(n)}_{BD}\nn
  +\alpha^2g(R(l,\partial_A)\partial_B,l)-\beta^2g(R(n,\partial_A)\partial_B,n)  +\lie_{N^T}K^{\tau}(\partial_A,\partial_B)\label{EOTK}
 \end{gather}

 \section{Inducing supertranslation via non expanding null surface}
 Before moving to the spherically symmetric case ket us explore the scenario considered in \cite{Ghosh:2020wjx}, where it was shown that supertranslation can be induced via a non expanding null surface. It is beyond our current understanding whether there can be any such physical process, since the stress energy tensor violated a classical energy condition. Let us try to see how a different choice of evolution vector leads to no supertranslation. The main reason we want to discuss this is to see that indeed an diffeomorphism on the cross sections is induced and it violates the basic assumptions that were made. In such a case the evolution of the divergence of the rotation one form is obtained with $X$ taken to be null i.e equal to $l$ and is given by,
 \begin{gather}
 \lie_X(div ~\omega)=\mathcal D^A\bigg(R(l,\partial_A)\bigg)-\bigg(\mathcal D^2\omega_B-\frac{~^2\mathcal R}{2} ~\omega_B\bigg)\bigg(\epsilon^{BC}\partial_Cg\bigg)-2\mathcal D^A\mathcal D^B\Omega~\mathcal D_A\mathcal D_Bg
 \end{gather}
 Note that since the expansion is zero one does not need to add a divergence part to $N^T$. The divergence free part is obtained by equating the right hand side of the above equation to zero. The addition of a divergence free part of $N^T$ however produces an extra diffeomorphism on the sphere. Clearly the condition that $q$ is Lie dragged along $X$, made in \cite{Ghosh:2020wjx}, is no more satisfied thus violating the assumptions.
 \section{Spherically symmetric DH}
 For the spherically symmetric case, all angular momentum multipole moments must be zero. Thus $\omega_A$ must be of the form $\omega_A=\partial_A \omega$. Moreover the evolution of these multipole moments must also be zero. To check what constraints this impose, first note that spherical symmetry of the $DH$ implies that $\Sigma^{X^\perp}_{AB}=0$, ~$\mathcal D_AK^{X^\perp}=0$ and $X^T=0$. Thus from the evolution equation it follows that,
 \begin{gather}
 \mathcal D_{[A}\beta~\mathcal D_{B]}K^{(n)}=0,~~~R(\tau,\partial_A)=\partial_A T
 \end{gather}
 The first of the above equation implies that $\beta$ is some function of $K^{(n)}$. Along with  the fact that $\mathcal D_AK^{X^\perp}=0$ it follows that $\mathcal D_AK^{(n)}=\mathcal D_A\beta=0$. Thus the evolution of the divergence of $\omega_A$ reduces to,
 \begin{gather}\label{ev_eq}
 \frac{d}{d\lambda}(div~\omega)+\frac{\dot R}{R}(div~\omega)=\mathcal D_A\mathcal D_B\big(q^{AC}q^{DE}K^{\tau}_{CE}\big)+\mathcal D^2\kappa_X+\mathcal D^2T.
 \end{gather}
 where $\lambda$ is a parameter along $X$. Further one can conclude that the shears in the two null directions are proportional to each other i.e $\Sigma_{AB}^{(l)}=\beta\Sigma_{AB}^{(n)}$. 
 To get an idea about what the quantities in the above equation might be such that it represents correctly the dynamical phase conjured, we will try to set up a metric in the neighbourhood of the DH by strategically implementing the assumptions made above. For simplicity we will assume the case for the collapse of null dust. Thus we will consider the Vaidya space-time as our seed metric which we will modify so as to represent a phase where a supertranslation is being induced.
 \subsection{Example}
 The metric in the neighbourhood of the spherically symmetric dynamical horizon when a supertranslation is being induced, can be studied by taking the intrinsic metric and then evolving it into the bulk. If the extrinsic curvature of the DH as an embedding in space-time is denoted by $\mathcal K({\partial_{a},\partial_b})$ then introducing a supertranslation amounts to modifying $\mathcal K(X,\partial_A)$. This is a 3+1 approach to the problem. In a 2+2 approach data on a cross-section of $\mathcal H$, is taken and the evolution along the two null directions normal to the surface is considered in order to obtain the metric in the neighbourhood. We will avoid both these approaches. Instead we will start with  an ansatz for the metric in the neighbourhood of $\mathcal H$ and derive conditions such that it is consistent with the assumptions made. The generic metric to study the effects of supertranslations for a spherically symmetric DH
 is given by the following form:
 \begin{equation}
 ds^2 = -fdv^{2}+2dvdr-2\partial_AC~dvd\tau^A 
+\mathscr R^2 d\theta^2+\mathscr R^2\sin^{2}\theta\,d\phi^2,\label{metric}
 \end{equation}
 \begin{gather}
 \underline{n}:=-dv,~~n=-\frac{\partial}{\partial r},~~~~~~\nn
 l:=\frac{\partial}{\partial v}-\bigg[\frac{f}{2}+\mathcal D^AC\mathcal D_AC\bigg]\frac{\partial}{\partial r}-\mathcal D^A C\partial_A\label{bases}
 \end{gather}
 where $f:=f(v,r,\theta,\phi),~\mathscr R:=\mathscr R(v,r,\theta,\phi),~C:=C(v,r,\theta,\phi)$. We want this to represent the phase where a supertranslation is being induced along with the increase of mass. So we clearly require the location and other structures of the horizon to remain same. Thus we demand that the spacetime be a minimal modification of the Vaidya spacetime so as to incorporate the non zero rotation form. First, the location of the horizon is assumed to be a level surface for a function of the coordinates $r$ and $v$. The requirement that the first order structure on the DH that is the intrinsic metric be unaltered, implies that the function $f$ must be such that $f|_{\mathcal H}=0$. Further since we require the dynamical horizon to be spherically symmetric, the vanishing of the non diagonal terms imply that $C$ must satisfy $C|_{\mathcal H}=0$. The condition on $\kappa_{X^\perp}$ used in eq. (\ref{evos}) then implies that $~\partial_rf|_{\mathcal H}=\frac{1}{2R}$, where $R$ is the areal radius of the horizon cross-sections. In order that the areal radius be corrected reflected in the metric we also must have $~\mathscr R|_{\mathcal H}=R$. The condition that expansion of the null normal $l$ be zero on the DH then implies that $\partial_v \mathscr R |_{\mathcal H}=0$. The evolution equation for areal radius $\partial_v\mathscr R-\beta\partial_rR=\dot R$ allows us to choose $\partial_rR=1$. This choice of metric gives the following quantities which are of direct interest to us as they need to be compared with the forms assumed during general considerations. The expression for the rotation one form and the surface gravity obtained using these choices are given by,
 \begin{gather}
 \omega_A=\frac{1}{2}\partial_A\partial_rC\bigg|_{\mathcal H},~~~~~\kappa_{X^\perp}=\frac{\partial_rf}{2}\bigg|_{\mathcal H}=\frac{1}{2R}=\frac{1}{4m(\lambda)},
 \end{gather}
 where  $m$ is the usual mass parameter of Vaidya spacetime. It is clear that $\omega_A$ is of the desired form and that $\omega=\frac{\partial_rC}{2}$. The extrinsic curvature evaluated on the the DH reads,
 \begin{gather}
 K_{AB}^{(l)}=0,~~~~~ K_{AB}^{(n)}=\frac{1}{R}~q_{AB}
 \end{gather}
Thus the condition is weaker than anticipated and will thus lead to easier set of equations for the evolution of the divergence of the rotation one form.  Let us now go over to the Einstein's equations. The metric when expanded in powers of $r-2m(v)$ is not expected to satisfy Einstein's equation, as there will be a mixing of orders. In particular, Einstein's equation on the DH will contain contribution from next to leading order term for $R$ and $C$. It can however be shown that with appropriate choices of the metric coefficients $G(l,n)$ on $\mathcal H$ can be held to zero. This is because of the fact that $T(l,n)$ is just a boundary data. This however will lead to an extra contribution to $G(l,l)$ apart from the usual matter flux term $\dot m$. Note that this kind of an argument is essential because neither do we have a knowledge of the stress energy tensor nor do we have a knowledge of how the metric in the neighbourhood will look like. The only information we have are about some quantities defined on the DH and the fact that the initial and final black holes must be stationary and isolated. This in contrast to asymptotic null infinity where supertranslation is a symmetry even in the dynamic phase, a universal Minkowski metric to work with, a knowledge of fall off conditions and the possibility to find the asymptotic form of the metric for any dynamic process that takes place in the bulk. In the case of a dynamical horizon we therefore must restrict our attention on the horizon structures available and try to fix relevant components of the stress energy tensor from these considerations alone. To see that this indeed can be done, let us recall the equation that determines whether a marginally trapped surface evolves to another marginally trapped surface. The consistency of eq.(\ref{MOTS}) with the assumption that $\beta=\dot m$ requires that the term $\mathcal D_A\omega\mathcal D^A\omega-\mathcal D^2\omega$ be somehow cancelled as this is the only term that is a function of the coordinates on the cross-section. One might remove this by choosing an appropriate $T(l,n)$. But note that this will imply that there will be a residual non zero $T(l,n)$ present at the end of the process, thus violating the condition that the final black hole is isolated with no flux. However if $T(l,l)$ is modified to get rid of this term, then this extra term in $T(l,l)$ will come with a multiplicative factor $\dot m$ which would ensure that $T(l,l)$ falls off to zero at the end of the process. Hence this will be the appropriate choice. Note that there enough freedom in the choice of the metric in the neighbourhood so as to accommodate this choice, again because of the fact that this a boundary data. Let us cross-check that this is indeed the case. Take $G(l,n)$ e.g. When restricted to the DH this yields, for the chosen metric,
\begin{gather}
G(l,n)\stackrel{\mathcal H}{=}
\omega^A\omega_A+\mathcal D^A\omega_A
+\frac{2\partial_r\partial_v\mathscr R}{\mathscr R}\bigg|_{\mathcal H}
\end{gather}
Therefore by appropriately choosing $\partial_r\partial_v\mathscr R|_{\mathcal H}$, $T(l,n)$ can be held to zero.
Similarly, $T(l,l)$ will contain a term of the form $\partial_v^2\mathscr R$ which when appropriately chosen gives the required choice for $T(l,l)$. To see this note that,
\begin{gather}
g(R(l,\partial_\theta)\partial_\theta,l)\stackrel{\mathcal H}{=}\bigg[\mathscr R~\partial_v^2\mathscr R+\partial_v\partial_\theta^2C+\frac{1}{2}\mathscr R\partial_vf~\partial_r\mathscr R\bigg]\bigg|_{\mathcal H}\nn
g(R(l,\partial_\phi)\partial_\phi,l)\stackrel{\mathcal H}{=}\bigg[\mathscr R\sin^2\theta~\partial_v^2\mathscr R+\partial_v\partial_\phi^2C+\sin\theta~\cos\theta~\partial_v\partial_\phi C+\frac{\sin^2\theta}{2}\mathscr  R~\partial_vf\partial_r\mathscr R\bigg]\bigg|_{\mathcal H}
\end{gather}
Thus,
\begin{gather}
G(l,l)\stackrel{\mathcal H}{=}\frac{1}{\mathscr R^2}\bigg[2\mathscr R~\partial_v^2\mathscr R+\mathscr R~\partial_vf~\partial_r\mathscr R+\mathscr D^2(\partial_vC)\bigg]\bigg|_{\mathcal H},
\end{gather}
where $\mathscr D$ is the covariant derivative on the unit sphere. These choices will determine how the metric in the neighbourhood will look like. As a final cross check we will see if $T(\tau,\partial_A)$ is indeed a total derivative. We will check this for $T(n,\partial_A)$. 
\begin{gather}
G(n,\partial_A)\stackrel{\mathcal H}{=}-\frac{1}{2\mathscr R^2}\bigg[\mathscr R^2\partial_A\partial_r^2C-2\partial_r\mathscr R ~\partial_A\mathscr R +2\mathscr R~\partial_A\partial_r\mathscr R\bigg]\bigg|_{\mathcal H}=-\frac{1}{2}\bigg[\partial_A\partial_r^2C+2\partial_A\partial_r\log \mathscr R\bigg]\bigg|_{\mathcal H}
\end{gather}

Such a result holds for $G(l,\partial_A)$ as well thus ensuring that $G(\tau,\partial_A)$ is indeed a total derivative. But the most important thing to note here is that contains an independent function $\partial_r^2C$ and is thus a data independent of those specified previously. With these assumption made the evolution of the divergence of the rotation one form is simply,
\begin{gather}\label{ev_eq}
\frac{d}{d\lambda}(div~\omega)+\frac{2\dot R}{R}(div~\omega)=\mathcal D^2T.
\end{gather}
It is clear from the above equation that for a spherically symmetric collapse (not only a spherically symmetric dynamical horizon) there is not supertranslation induced. This because spherical symmetry will imply that therms like $T(\partial_A,\tau)$ are zero. 

 As a final consistency check we will investigate the evolution equation for the transverse extrinsic curvature. Note that for the spherically symmetric case, both $\alpha$ and $\beta$ are constants on the cross-sections. Thus the right hand side of eq .(\ref{EOTK}) simplifies to,
\begin{gather}
\bigg(\kappa_{X^\perp}-\frac{1}{2}\frac{\nabla_{X^\perp}\beta}{\beta}\bigg)K^{X^\perp}_{AB}+\frac{1}{2}\bigg(\frac{\nabla_{X^\perp}\beta}{\beta}\bigg)K^{\tau}_{AB}-2\beta\mathcal D_A\mathcal D_B\omega+\beta^2K^{(n)}_{AC}q^{CD}K^{(n)}_{DB}\nn
+g(R(l,\partial_A)\partial_B,l)-\beta^2g(R(n,\partial_A)\partial_B,n)
\end{gather}
The expressions for the terms calculated from the metric eq. (\ref{metric}) and basis vectors eq. (\ref{bases}), when restricted to the horizon gives the following,
\begin{gather}
g(R(l,\partial_A)\partial_B,l)=\mathcal D_A\mathcal D_B\partial_vC\bigg|_{\mathcal H}+\frac{q_{AB}}{\mathscr R}\bigg(\partial_v^2\mathscr R+\frac{\partial_vf~\partial_r\mathscr R-\partial_rf~\partial_v\mathscr R}{2}\bigg)\bigg|_{\mathcal H}\nn
g(R(n,\partial_A)\partial_B,n)=\frac{q_{AB}}{\mathscr R}\partial_r^2\mathscr R\bigg|_{\mathcal H}\nn
K^{(l)}_{AB}=0,~~~K^{(n)}_{AB}=\frac{1}{\mathscr R}q_{AB}\bigg|_{\mathcal H},~~~K^{X^\perp}_{AB}=-\frac{\beta}{\mathscr R}q_{AB}\bigg|_{\mathcal H},~~~K^{\tau}_{AB}=\frac{\beta}{\mathscr R}q_{AB}\bigg|_{\mathcal H},~~~\kappa_{X^\perp}=\frac{1}{2R}
\end{gather}
Now, note that $\lie_XC|_{\mathcal H}=0$. Thus $\partial_vC\stackrel{\mathcal H}{=}\beta \partial_rC$.
Further note that $\lie_Xf=0$ which implies $\partial_vf=\beta~\partial_rf$. Using these constraints and those spelled out before, the right hand side of eq.(\ref{EOTK}) gives,
\begin{gather}
\frac{\partial_\lambda\beta}{R}q_{AB}+\frac{\beta^2}{R^2}q_{AB}+\frac{\partial_v^2\mathscr R-\beta^2\partial_r^2\mathscr R}{\mathscr R}q_{AB}\bigg|_{\mathcal H}
\end{gather}
The left hand side is the derivative of $K^{\tau}_{AB}$, which gives,
\begin{gather}
\frac{\dot\beta}{R}q_{AB}-\frac{\beta\dot R}{R^2}q_{AB}+\frac{2\beta^2}{R^2}q_{AB}
\end{gather}
Equating the left and right hand sides we have the constraint $\bigg(\partial_v^2\mathscr R-\beta^2\partial_r^2\mathscr R\bigg)\bigg|_{\mathcal H}=0$. Note that all these conditions, some imposed and some arrived at from consistency requirements, are to be seen to hold on the cross-sections $S_{\mathcal H}$ of $\mathcal H$.

\begin{figure}[h]
 	\centering
 	\includegraphics[width=0.35\linewidth]{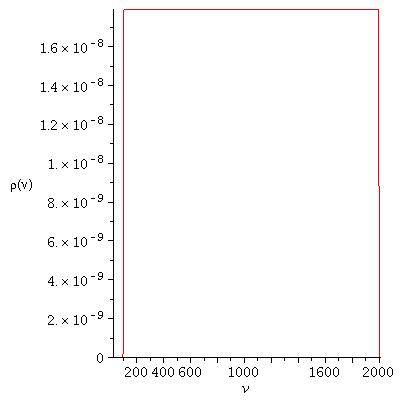}
 	\caption{The figure shows form of the function $\rho(v)$ in equation \eqref{rho_expression}.}
 	\label{fig1}
 \end{figure}

\begin{figure}[h]
 	\begin{subfigure}{.4\textwidth}
 		\centering
 		\includegraphics[width=\linewidth]{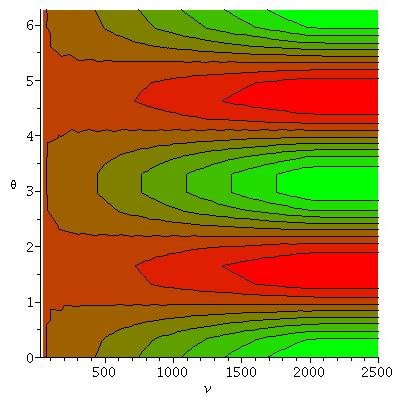}
 		\caption{}
 	\end{subfigure}
 	\begin{subfigure}{.4\textwidth}
 		\centering
 		\includegraphics[width=\linewidth]{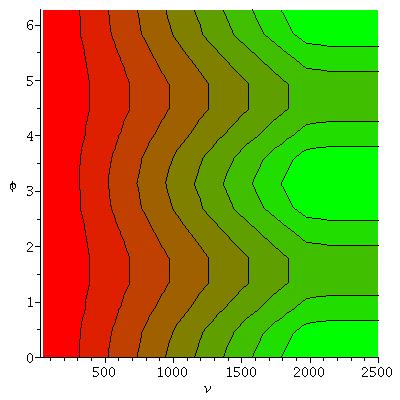}
 		\caption{}
 	\end{subfigure}
 	\caption{The figures show the variation of $div~ \omega$ with $v$.
 		Note that the variation of $div ~\omega$ begins from $div ~\omega=0$ at
 		around $v=100$ in harmony with variation of the function $T(v,\theta,\phi)$
 		in figure $1$. The variation stops at around $v=2010$ 
 		with the function $T(v,\theta,\phi)$. The graph
 		in (a) is for a fixed $\phi=\pi/4$ while (b) is for $\theta=\pi/10$.}
 	\label{fig2}
 \end{figure}

 Having obtained the necessary constraints and shown that the metric ansatz is indeed the correct one, let us consider a simple graphical illustration of our claim that the divergence of $\omega_A$ evolves. We shall solve the equation 
 \eqref{ev_eq} for this case of spherical symmetry. Rewriting the equation in terms of the covariant derivative on the unit sphere, we have,
 \begin{equation}
 \frac{d}{d\lambda} (\mathscr D^A\omega_A)=\mathscr D^{2} T.
 \end{equation}
 In the next discussion we will identify $\lambda$ with $v$. For our purposes, we shall use the form of $T(v,\theta, \phi)$ such that it has support only
 during the time the matter falls and vanishes otherwise. One such form of $T(v,\theta,\phi)$
 is to assume a separable form $T(v,\theta, \phi)=\bar{T}(\theta, \phi)\rho(v)$, where we consider the
 $\rho(v)$ to be:
 \begin{equation}\label{rho_expression}
 \rho(v)=\frac{3a}{4\pi (y-x)(2x^{2}+2y^{2}+2xy+3M^{2})}\left[\erf\left(\frac{v-x}{M}\right)-
 \erf\left(\frac{v-y}{M}\right)\right],
 \end{equation}
 where $\erf(\alpha)$ is the error function
 and for the present computation, we shall assume $M=1$, $x=100$, $y=2000$, $M=1$ and $a=600 M$. 
 The form of the function $\rho(v)$ is plotted in Fig. \eqref{fig1}.
 For solving this differential equation, we shall use the boundary condition that 
 the function $(\mathscr D^A\omega_A)$ vanishes at the beginning of the process, at $v=100$.
 For $\bar{T}(\theta, \phi)=\sin^{2}\theta\sin^{2}\phi$, the equation can be solved, leading to the 
 variation of $(\mathscr D^A\omega_A)$ with $v$. For a fixed set of values of 
 $\theta$ and $\phi$, this variation is plotted as a function of $v$ in Figure \eqref{fig2}.

 \section{Conclusion}
 In this paper we have defined a notion, when we might say that a supertranslation is being induced during a dynamical evolution. The definition relies on the concept of choosing a preferred foliation for an Isolated horizon. If during the dynamical evolution from one isolated black hole state to another an initial black hole, which is preferably foliated, evolves to another which is not, then we would conclude that a supertranslation has been induced. This is in agreement with the fact that a supertranslation acting on any isolated horizon data essentially changes its foliation. Thus a natural way to conclude whether a supertranslation has been induced is to check whether the condition for choosing preferred foliation is preserved with time.
 
 There are various ways to define preferred slices. In this context we have worked with the condition that the divergence of the rotation one form is zero. We show that in general this condition is not preserved in the course of evolution along the dynamical horizon. However by a judicious choice of the divergence free part of the shift vector, on the DH, one can set the derivative of the divergence to zero. The existence of such a choice however depends on the existence of a global solution to a second order partial differential equation on the sphere. But since this implies a modification of the shift vector an extra diffeomorphism on the two sphere cross -section are automatically induced. The modification of the shift is also reflected in the evolution equation for the multipole moments as it is not completely independent of the divergence free part of the shift vector. In cases where the shift vector is constrained by symmetry requirements, on the DH, there is not much freedom in making this choice. Thus in general a supertranslation will be induced in the course of evolution. 
 
 It is observed that the divergence part of the $T(\tau,\partial_A)$ component of the stress energy tensor is a crucial data which decides whether a supertranslation is induced. This fact is clearly visible, when simplified version of the evolution equation, tailored for handling spherically symmetric DH's, are studied in detail. The exploration of such spherically symmetric DH,  evolving due to a collapse of null dust, but now with an additional supertranslation being induced, gives us further insights. We assume an ansatz for the metric in the neighborhood of the DH, which is motivated by the fact that it should be a minimal modification of the Vaidya space-time. We then go on to impose conditions on the values of the metric coefficients and its derivatives on the DH such that it reflects the process. This is checked by comparing various quantities calculated from the metric with those derived from general considerations. Consequently, we can partially fix the flux of stress energy tensor required to carry out such a process. The main conclusion that one draws from this is that the stress energy tensor violates the dominant energy condition in agreement with \cite{Haco:2018ske,Ghosh:2020wjx}.
 
The case where a modification of the shift vector might render the overall change in the divergenece of the rotation one form trivial is howver beyond the scope of the current work as it would require a more rigorous numerical approach. Likewise a more robust test of the proposals requires looking into more general dynamical process which are within the realm of numerical relativity and is beyond the scope of the current work.

\acknowledgements
The author AC  supported through the DAE-BRNS project number $58/14/25/2019$-BRNS
and the DST-MATRICS scheme of government of India through grant number MTR$/2019/000916$. A.G acknowledges the support through a grant from the NSF of China with Grant No: 11947301.

\appendix
\section{Notations and conventions}\label{N&C}
 The covariant derivative $\nabla: T\mathcal M \otimes T\mathcal M\rightarrow T\mathcal M$  will be denoted by $\nabla_WZ$ where $W,Z~\in ~T\mathcal M$. If $\mathcal S$ is an immersed submanifold then the tangent space at any point $x\in\mathcal S$ can be decomposed as $T_x\mathcal M=T_x\mathcal S\oplus T_x^{\perp}\mathcal S$. The covariant derivative on $\mathcal S$ denoted by $\mathcal D_XY$, where $X,Y~\in T\mathcal S$ is related to the covariant derivative $\nabla$ via the Gauss decomposition,
\begin{gather}
\nabla_XY=\mathcal D_XY+K(X,Y),
\end{gather}
where $K(X,Y)$ is the extrinsic curvature. Denoting the connection in the  normal bundle as $\nabla^{\perp}_XN^\perp$, where $X\in T\mathcal S$ and $N^\perp\in T^\perp\mathcal S $. the shape operator $W_{N^\perp}(X)$ can be defined as,
\begin{gather}
\nabla_XN^\perp=\nabla_X^\perp N^\perp-W_{N^\perp}(X).
\end{gather} 
The shape operator and the extrinsic curvature are therefore related by,
\begin{gather}
g(W_{N^\perp}(X),Y)=g(N^\perp,K(X,Y)),
\end{gather}
where $X,Y\in T\mathcal S$ and $N^\perp\in T^\perp\mathcal S$. The Riemann tensor is defined as,
\begin{gather}
R(W,U)V\equiv[\nabla_W,\nabla_U]V-\nabla_{[W,U]}V
\end{gather}
Similarly one can define an intrinsic Riemann tensor as,
\begin{gather}
\mathcal R(X,Y)Z\equiv[D_X,D_Y]Z-D_{[X,Y]}Z
\end{gather}
Using these definitions  the equations of Gauss and Codazzi can be written down. Let $X,Y,Z,W\in T\mathcal S$ and $N^\perp\in T^\perp\mathcal S$. Then the Gauss equation is given as,
\begin{gather}
g(R(X,Y)Z,W)=g(\mathcal R(X,Y)Z,W)-g(K(X,Z),K(Y,W))+g(K(X,W),K(Y,Z)),\label{Gauss}
\end{gather}
and the Codazzi equation as,
\begin{gather}
g(R(X,Y)N^\perp,Z)=g((\nabla_YK)(X,Z),N^\perp)-g((\nabla_XK)(Y,Z),N^\perp)\label{Codazzi}
\end{gather}
%
%
%
%
%

\end{document}